\documentclass[reprint,amsmath,amssymb,superscriptaddress]{revtex4-1}

\usepackage{graphicx}
\usepackage{dcolumn}
\usepackage{bm}
\usepackage{color}
\begin{document}

% Use the \preprint command to place your local institutional report
% number in the upper righthand corner of the title page in preprint mode.
% Multiple \preprint commands are allowed.
% Use the 'preprintnumbers' class option to override journal defaults
% to display numbers if necessary
%\preprint{}

%Title of paper
%\title{Large tunnel magnetoresistance in (111)-oriented junctions with SrTiO$_3$ barriers: Coherent tunneling of half-metallic states driven by the band-folding effect}
\title{Band-folding-driven high tunnel magnetoresistance ratios in (111)-oriented junctions with SrTiO$_3$ barriers}

% repeat the \author .. \affiliation  etc. as needed
% \email, \thanks, \homepage, \altaffiliation all apply to the current
% author. Explanatory text should go in the []'s, actual e-mail
% address or url should go in the {}'s for \email and \homepage.
% Please use the appropriate macro foreach each type of information

% \affiliation command applies to all authors since the last
% \affiliation command. The \affiliation command should follow the
% other information
% \affiliation can be followed by \email, \homepage, \thanks as well.
\author{Keisuke Masuda}
\email{MASUDA.Keisuke@nims.go.jp}
\affiliation{Research Center for Magnetic and Spintronic Materials, National Institute for Materials Science (NIMS), Tsukuba 305-0047, Japan}
\author{Hiroyoshi Itoh}
\affiliation{Department of Pure and Applied Physics, Kansai University, Suita 564-8680, Japan}
\affiliation{Center for Spintronics Research Network, Osaka University, Toyonaka 560-8531, Japan}
\author{Yoshiaki Sonobe}
\affiliation{Research Organization for Nano $\&$ Life Innovation, Waseda University, Shinjuku 162-0041, Japan}
\author{Hiroaki Sukegawa}
\affiliation{Research Center for Magnetic and Spintronic Materials, National Institute for Materials Science (NIMS), Tsukuba 305-0047, Japan}

\author{Seiji Mitani}
\affiliation{Research Center for Magnetic and Spintronic Materials, National Institute for Materials Science (NIMS), Tsukuba 305-0047, Japan}
\affiliation{Graduate School of Science and Technology, University of Tsukuba, Tsukuba 305-8577, Japan}

\author{Yoshio Miura}
\affiliation{Research Center for Magnetic and Spintronic Materials, National Institute for Materials Science (NIMS), Tsukuba 305-0047, Japan}
\affiliation{Center for Spintronics Research Network, Osaka University, Toyonaka 560-8531, Japan}
%\email[]{Your e-mail address}
%\homepage[]{Your web page}
%\thanks{}
%\altaffiliation{}
%\affiliation{}

%Collaboration name if desired (requires use of superscriptaddress
%option in \documentclass). \noaffiliation is required (may also be
%used with the \author command).
%\collaboration can be followed by \email, \homepage, \thanks as well.
%\collaboration{}
%\noaffiliation

\date{\today}

\begin{abstract}
We theoretically study the tunnel magnetoresistance (TMR) effect in (111)-oriented magnetic tunnel junctions (MTJs) with SrTiO$_{3}$ barriers, Co/SrTiO$_{3}$/Co(111) and Ni/SrTiO$_{3}$/Ni(111). Our analysis combining the first-principles calculation and the Landauer formula shows that the Co-based MTJ has a high TMR ratio over 500\%, while the Ni-based MTJ has a smaller value (290\%). Since the in-plane lattice periodicity of SrTiO$_{3}$ is about twice that of the primitive cell of fcc Co (Ni), the original bands of Co (Ni) are folded in the $k_x$-$k_y$ plane corresponding to the $ab$ plane of the MTJ supercell. We find that this band folding gives a half-metallic band structure in the $\Lambda_1$ state of Co (Ni) and the coherent tunneling of such a half-metallic $\Lambda_1$ state yields a high TMR ratio. We also reveal that the difference in the TMR ratio between the Co- and Ni-based MTJs can be understood by different $s$-orbital weights in the $\Lambda_1$ band at the Fermi level.
\end{abstract}

% insert suggested PACS numbers in braces on next line
% insert suggested keywords - APS authors don't need to do this
%\keywords{}

%\maketitle must follow title, authors, abstract, \pacs, and \keywords
\maketitle

% body of paper here - Use proper section commands
% References should be done using the \cite, \ref, and \label commands
\section{\label{introduction} introduction}
The tunnel magnetoresistance (TMR) effect is essential not only for applications to magnetic sensors and memories but also for deepening our understanding of spin-dependent electron transport. A series of studies on Fe/MgO/Fe(001) magnetic tunnel junctions (MTJs) \cite{2001Butler-PRB,2001Mathon-PRB,2004Yuasa-NatMat,2004Parkin-NatMat} has established the so-called coherent tunneling mechanism, which explains the TMR effect by bulk band structures of bcc Fe and MgO. In the $\Delta$ line of the Brillouin zone corresponding to the [001] direction, MgO has the slowest-decaying evanescent state with $\Delta_1$ symmetry within the band gap, allowing the major contribution of the $\Delta_1$ wave function to the transmission. Since bcc Fe has a half-metallic band structure in the $\Delta_1$ state, the majority-spin $\Delta_1$ wave function can mainly tunnel through MgO, leading to a giant TMR effect. Because of the successful observation of high TMR ratios \cite{2004Yuasa-NatMat,2004Parkin-NatMat}, this mechanism has been widely accepted and bcc(001)-oriented MTJs with MgO barriers have been mainly studied from both experimental and theoretical points of view.

In contrast, our recent studies \cite{2020Masuda-PRB,2021Masuda-PRB} have focused on unconventional fcc(111)-oriented MTJs with the stacking direction parallel to [111] directions of both the fcc ferromagnetic electrode and the insulator barrier. These MTJs are advantageous for obtaining large perpendicular magnetic anisotropy (PMA), which is another requirement in addition to high TMR ratios for the application to magnetic random access memories. There are many fcc ferromagnetic materials with large magnetic anisotropy along their [111] directions. Moreover, the (111) plane of the fcc structure is the closed-packed plane and has the lowest surface energy, indicating that (111)-oriented MTJs are compatible with fcc ferromagnetic electrodes. Thus, we have investigated the potential of such MTJs in the TMR effect on the basis of the first-principles calculation. We have shown that several (111)-oriented MTJs with Co-based ferromagnetic electrodes and MgO barriers have high TMR ratios \cite{2020Masuda-PRB,2021Masuda-PRB}, which originate from the interfacial resonant tunneling, in contrast to the conventional coherent tunneling of bulk electronic states in ferromagnetic electrodes.

Although such a mechanism of high TMR ratios is physically significant, the interfacial resonant tunneling might be sensitive to atomic configurations at interfaces of MTJs. Moreover, the application of bias voltages tends to suppress the interfacial resonant tunneling, since the energy level of the interfacial state is shifted oppositely in the two interfaces. These motivate us to find other (111)-oriented MTJs with robustly high TMR ratios.

\begin{figure}[b]
\includegraphics[width=8.7cm]{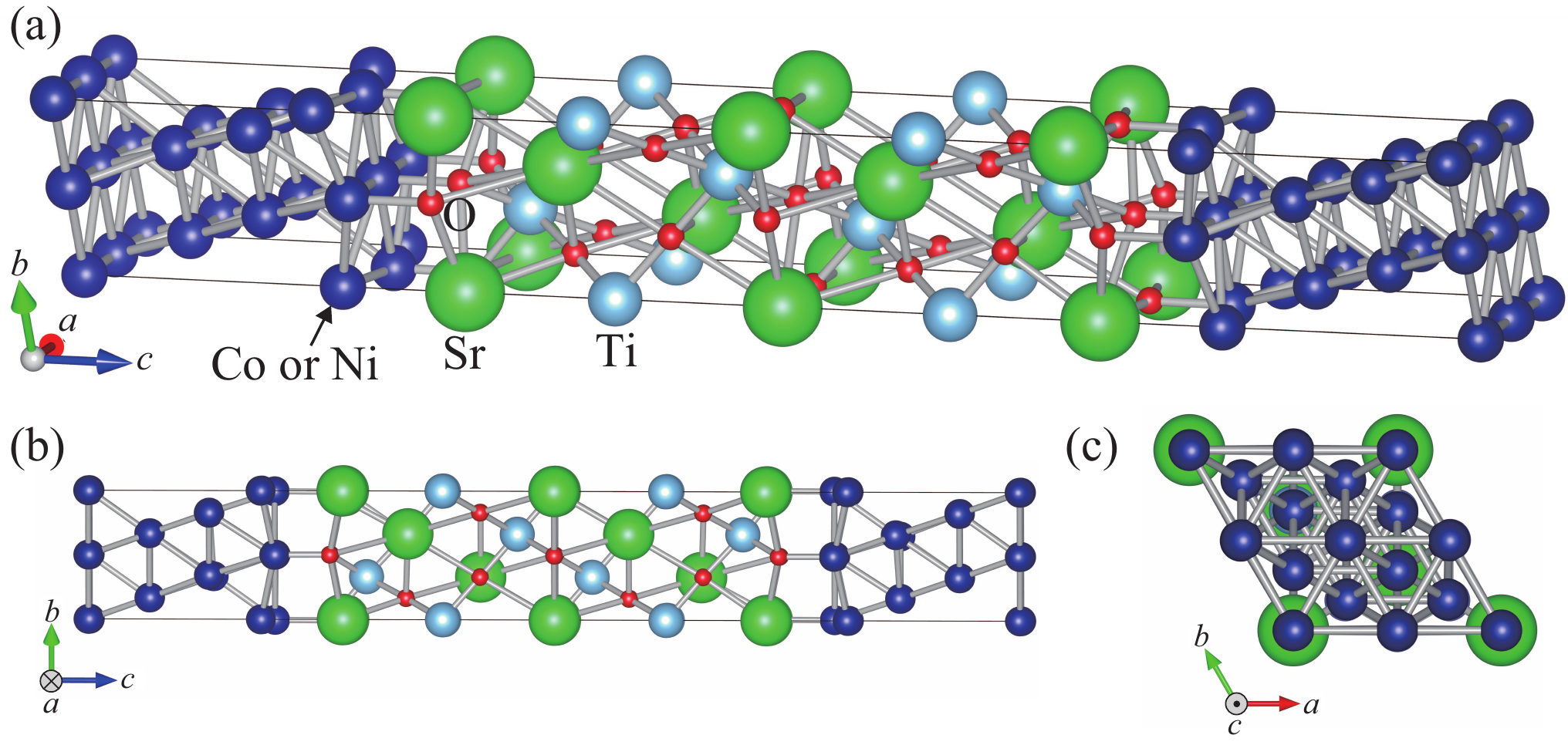}
\caption{\label{supercell} Supercell of X/SrTiO$_{3}$/X(111) (X=Co or Ni). (a) Three-dimensional view. (b) Side view from the $a$-axis and (c) top view from the $c$-axis directions.}
\end{figure}
\begin{figure}[b]
\includegraphics[width=8.2cm]{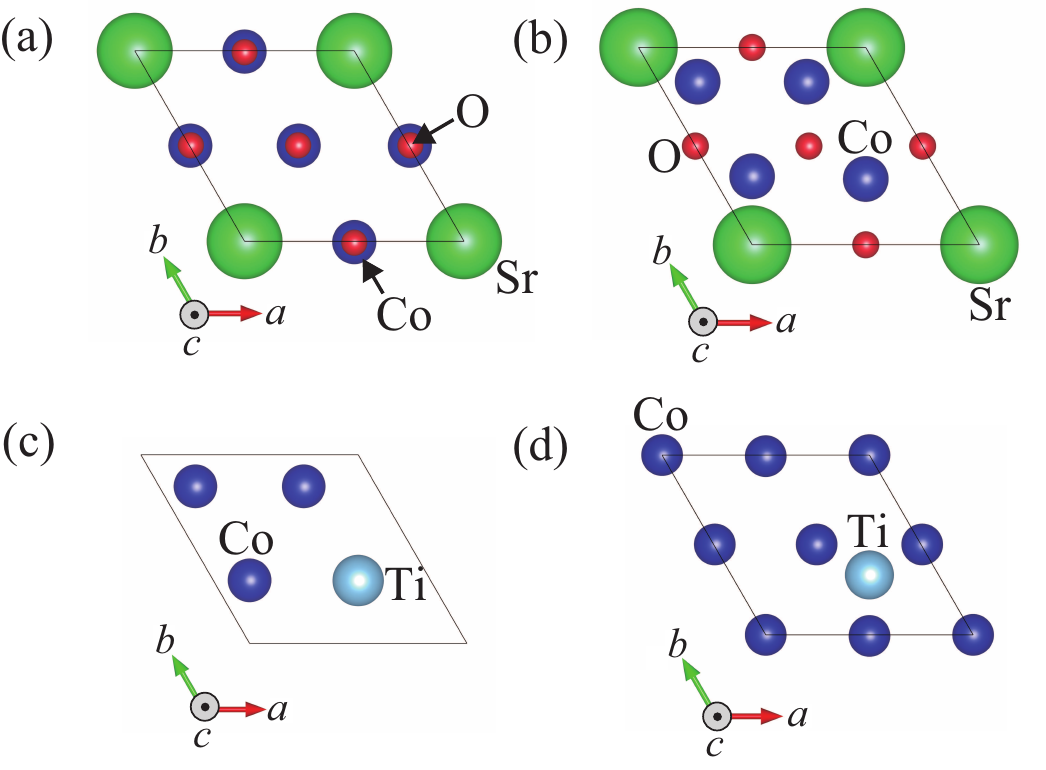}
\caption{\label{interface} Top view of different interfacial atomic configurations in Co/SrTiO$_{3}$/Co(111). (a),(b) SrO-terminated interfaces with (a) Sr and O on top of Co and (b) Sr and O on hollow sites. (c),(d) Ti-terminated interfaces with (c) Ti on top of Co and (d) Ti on hollow sites.}
\end{figure}
In this work, we consider (111)-oriented MTJs with SrTiO$_3$ tunnel barriers. Historically, SrTiO$_3$ has been recognized as an important material for tunnel barriers. Comparative experimental studies on Co/$X$/La$_{0.7}$Sr$_{0.3}$MnO$_3$ ($X$\,=\,SrTiO$_3$, Al$_2$O$_3$, and Ce$_{0.69}$La$_{0.31}$O$_{1.845}$) clarified that the spin polarization of effective tunneling electrons but not that of ferromagnets plays a crucial role in the TMR effect \cite{1999Teresa-PRL,1999Teresa-Science}. Moreover, a high TMR ratio was predicted theoretically in bcc(001)-oriented Co/SrTiO$_3$/Co(001) \cite{2005Velev-PRL}. Although experiments on such (001)-oriented MTJs have not succeeded in achieving high TMR ratios, unconventional (111)-oriented MTJs with SrTiO$_3$ barriers may open a pathway for high TMR ratios. We thus focus on fcc(111)-oriented MTJs, Co/SrTiO$_3$/Co(111) and Ni/SrTiO$_3$/Ni(111) (Fig. \ref{supercell}).

Our first-principles-based transport calculation demonstrates that the Co- and Ni-based MTJs show relatively high TMR ratios of 534 and 290\%, respectively. We also reveal that the high TMR ratios can be explained by the coherent tunneling of electronic states of bulk ferromagnets, meaning that the obtained TMR ratios are more robust against interfacial imperfections and bias voltage than those driven by the interfacial resonant tunneling. The simple fcc Co and Ni given by their primitive unit cells have no $\Lambda_1$ bands crossing the Fermi level in the high-symmetry line $\Lambda$ corresponding to the [111] direction. We show however that these ferromagnets have a half-metallic band structure in the $\Lambda_1$ state when attached to SrTiO$_{3}$, because the in-plane periodicity of SrTiO$_3$ is about twice that of fcc Co (or Ni) and the original band structure of Co (or Ni) is folded in the $k_x$-$k_y$ plane. This is a kind of ``band-folding effect'' found and studied in (001)-oriented MTJs with spinel-oxide tunnel barriers \cite{2012Miura-PRB,2012Sukegawa-PRB,2017Masuda-PRB,2020Nawa-PRB,2021Nawa-PRApplied}. Finally, we address the difference in the TMR ratio between the Co- and Ni-based MTJs and clarify that this comes from different $s$-orbital weights in the $\Lambda_1$ band at the Fermi level.

\section{\label{model-method} model and method}
We first considered supercells of Co/SrTiO$_3$/Co(111) and Ni/SrTiO$_3$/Ni(111) (Fig. \ref{supercell}), in which fcc Co (or Ni) and SrTiO$_3$ are stacked along their [111] directions. These supercells have the hexagonal close-packed structure given by the primitive translation vectors, ${\bm a}_1=a(1,0,0)$, ${\bm a}_2=a(-1/2,\sqrt{3}/2,0)$, and ${\bm a}_3=(0,0,c)$, where $a=\sqrt{2}\,a_{\rm fcc}$ with $a_{\rm fcc}$ being the lattice constant of fcc Co (or Ni) and $c$ is the length of the supercell. We used $a_{\rm fcc}=3.52\,{\rm \AA}$ for both the supercells and fitted SrTiO$_3$ to fcc Co (or Ni) in the $ab$ plane. The length $c$ was determined by the structure optimization mentioned below. The supercell includes 13 monolayers (ML) of SrTiO$_3$ and 7 ML of Co (or Ni). The thickness of SrTiO$_3$ layers is approximately 1.9\,nm \cite{remark_barrier-thickness}, which is a typical barrier thickness (1--2\,nm) used in MTJs.

\begin{table}
\caption{\label{tab0}
Formation energy divided by the cell volume $E_{\rm form}/V$ in each Co/SrTiO$_3$/Co(111) supercell.
}
\begin{ruledtabular}
\begin{tabular}{cccc}
 & interfacial atomic configuration & $E_{\rm form}/V\,\, ({\rm eV/\AA^3})$ & \\
\colrule
 & Co-SrO (on-top) & -1.899$\times$10$^{-1}$ & \\
 & Co-SrO (hollow) & -1.878$\times$10$^{-1}$ & \\
 & Co-Ti  (hollow) & -1.513$\times$10$^{-1}$ & \\
 & Co-Ti  (on-top) & -1.506$\times$10$^{-1}$ & \\
\end{tabular}
\end{ruledtabular}
\end{table}
Note here that there are four possible candidates of the interfacial atomic configuration of the supercell as shown in Fig. \ref{interface}. After preparing supercells for all the cases, atomic positions in each supercell were relaxed along the $c$ direction and the formation energy of each supercell was calculated. To determine the energetically favored supercell, we compared formation energies of Co/SrTiO$_3$/Co(111) supercells with different interfacial atomic configurations [Figs. \ref{interface}(a)--\ref{interface}(d)]. The formation energy for each supercell is expressed as
\begin{equation}
E_{\rm form}=E_{\rm tot}-\sum_{i} N_i \mu_i,\label{formene}
\end{equation}
where $E_{\rm tot}$ is the total energy of the optimized supercell with each interfacial atomic configuration, $N_i$ is the number of atoms of the element $i$ and $\mu_i$ is its chemical potential. In the present work, we used $\mu_{\rm Co}$, $\mu_{\rm Sr}$, $\mu_{\rm Ti}$, and $\mu_{\rm O}$ derived from energies of hcp Co, fcc Sr, hcp Ti, and O$_2$ molecules. Table \ref{tab0} shows obtained formation energies. We find that the supercell with the SrO-terminated interface with Sr and O on top of Co [Fig. \ref{interface}(a)] has the lowest value of $E_{\rm form}/V$. Therefore, this supercell (Fig. \ref{supercell}) was selected for the calculation of the TMR ratio. All the structural optimizations were performed using the first-principles calculation based on the density-functional theory (DFT) implemented in the Vienna {\it ab initio} simulation program ({\scriptsize VASP}) \cite{1996Kresse-PRB}. We adopted the generalized gradient approximation (GGA) \cite{1996Perdew-PRL} for the exchange-correlation energy and used the projected augmented wave (PAW) pseudopotential \cite{1994Bloechl-PRB,1999Kresse-PRB} to treat the effect of core electrons properly. A cutoff energy of 500\,eV was employed and the Brillouin-zone integration was performed with $13\times13\times1$ {\bf k} points. More details of the structural optimization are mentioned in our previous work \cite{2017Masuda-PRB}.

We calculated the TMR ratios on the basis of the ballistic transport theory. The Landauer formula was used in ballistic transport calculations with the first-principles DFT method, which is implemented in the {\scriptsize PWCOND} code \cite{2004Smogunov-PRB} in the {\scriptsize QUANTUM ESPRESSO} package \cite{Baroni}. We first constructed the quantum open system by attaching the left and right semi-infinite electrodes of fcc Co (Ni) to the supercell Co/SrTiO$_3$/Co (Ni/SrTiO$_3$/Ni). Then, the self-consistent potential of the quantum open system was obtained by the first-principles calculation, where the GGA and the ultrasoft pseudopotentials \cite{2014DalCorso_CMS} were used. The cutoff energies for the wave functions and the charge density were fixed to 58 and 580\,Ry, respectively, and 13$\times$13$\times$1 {\bf k} points were used for the Brillouin-zone integration. Since the quantum open system has the translational symmetry in the $ab$ plane, the scattering state can be classified by an in-plane wave vector ${\bf k}_{\parallel}=(k_x,k_y)$, where the $x$ axis was set to be parallel to the $a$ axis. Here, $(x,y)$ and $(k_x,k_y)$ are given in the Cartesian coordinates. For each ${\bf k}_{\parallel}$ and spin index, we solved the scattering equation derived under the condition that the wave function and its derivative of the supercell are connected to those of the electrodes \cite{1999Choi-PRB,2004Smogunov-PRB}. From the obtained transmittance, we calculated the conductance using the Landauer formula. These calculations for both parallel (P) and antiparallel (AP) magnetization states of electrodes provide the following wave-vector-resolved conductances: $G_{{\rm P},\uparrow}({\bf k}_{\parallel})$, $G_{{\rm P},\downarrow}({\bf k}_{\parallel})$, $G_{{\rm AP},\uparrow}({\bf k}_{\parallel})$, and $G_{{\rm AP},\downarrow}({\bf k}_{\parallel})$, where $\uparrow$ ($\downarrow$) indicates the up-spin (down-spin) channel. In this work, the up-spin (down-spin) channel is defined as the majority-spin (minority-spin) channel of the left electrode in both the parallel and antiparallel magnetization states. We calculated the averaged conductances as, e.g., $G_{{\rm P},\uparrow}=\sum_{{\bf k}_\parallel}G_{{\rm P},\uparrow}({\bf k}_\parallel)/N$, where $N$ is the sampling number of ${\bf k}_{\parallel}$ points. After confirming good convergence of the conductances and TMR ratio, $N$ was set to 150$\times$150\,=\,2250. Using the averaged conductances, we calculated the TMR ratio given by the optimistic definition, i.e., ${\rm TMR\,\, ratio}\,(\%)=100\times(G_{\rm P}-G_{\rm AP})/G_{\rm AP}$, where $G_{\rm P(AP)}=G_{{\rm P(AP)},\uparrow}+G_{{\rm P(AP)},\downarrow}$.

\section{\label{resultsdiscussion} results and discussion}
\begin{figure*}[t]
\begin{center}
\includegraphics[width=17.0cm]{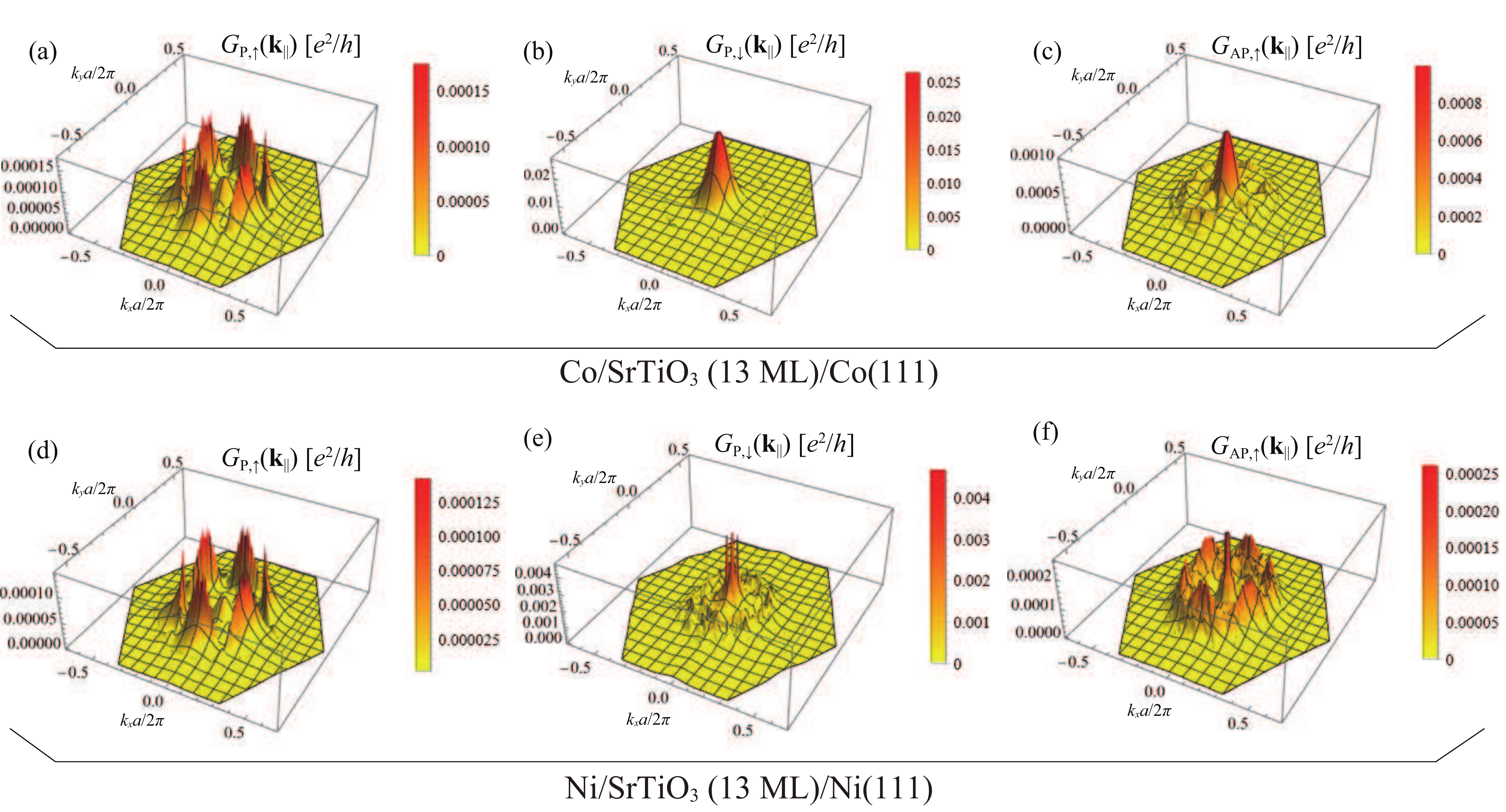}
\caption{\label{kpara} ${\bf k}_{\parallel}$-dependent conductances in Co/SrTiO$_3$\,(13\,ML)/Co(111) [(a)--(c)] and Ni/SrTiO$_3$\,(13\,ML)/Ni(111) [(d)--(f)], where ${\bf k}_{\parallel}=(k_x,k_y)$ is given in the Cartesian coordinates. (a),(d) Up-spin conductances $G_{{\rm P},\uparrow}({\bf k}_{\parallel})$ and (b),(e) down-spin conductances $G_{{\rm P},\downarrow}({\bf k}_{\parallel})$ in the parallel magnetization configurations. (c),(f) Up-spin conductances $G_{{\rm AP},\uparrow}({\bf k}_{\parallel})$ in the antiparallel magnetization configurations.}
\end{center}
\end{figure*}
\begin{figure*}[t]
\begin{center}
\includegraphics[width=17.0cm]{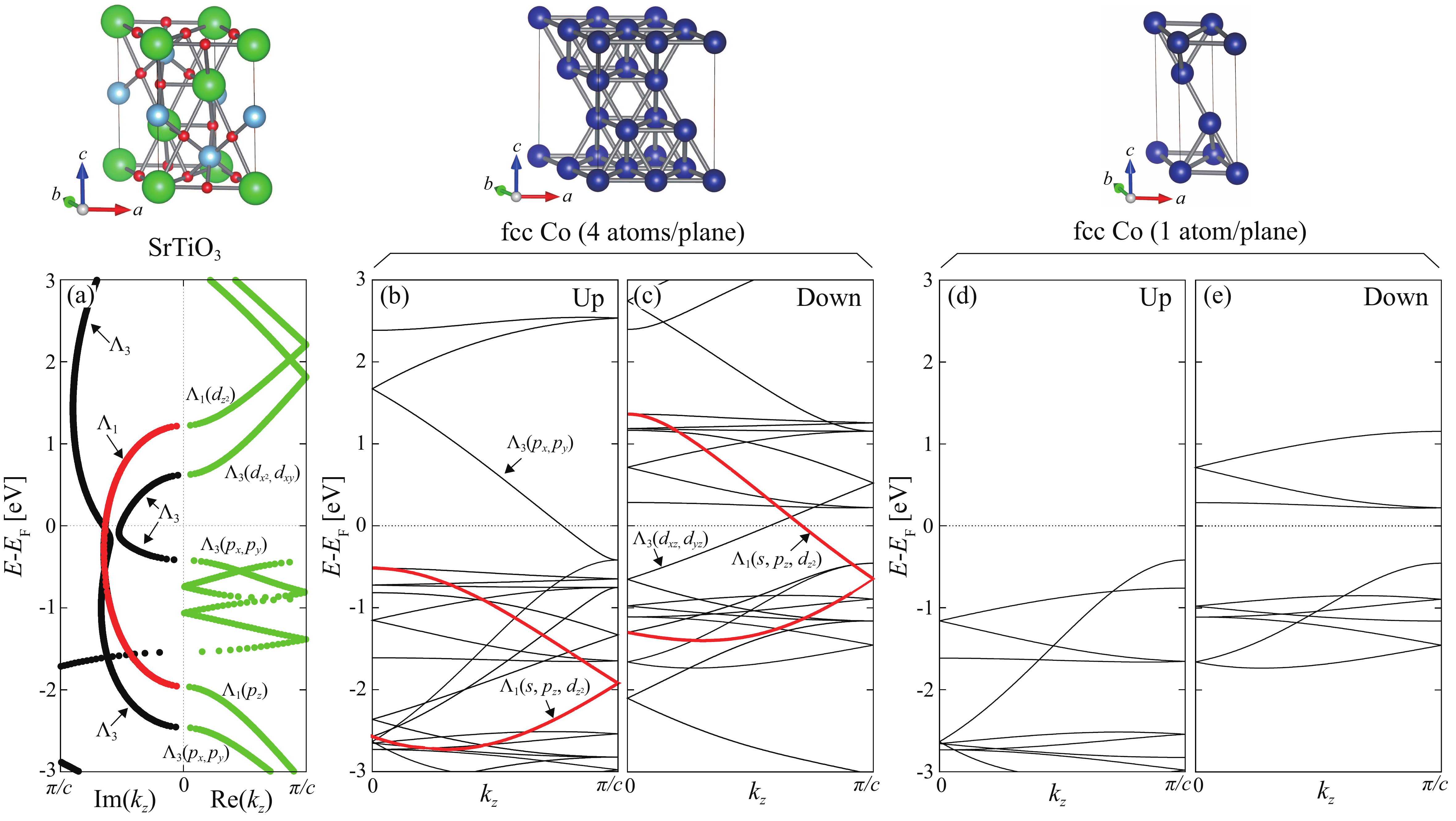}
\caption{\label{band-mixed} (a) Real and complex band structures along the $\Lambda$ line of SrTiO$_3$. (b) Up-spin and (c) down-spin band structures along the $\Lambda$ line of fcc Co calculated for the unit cell with four atoms in each plane. In (a)--(c), the irreducible representation and atomic orbitals contributing dominantly to each band are indicated, where $d_{3z^2-r^2}$ and $d_{x^2-y^2}$ are abbreviated as $d_{z^2}$ and $d_{x^2}$, respectively. (d),(e) The same as (b),(c) but for the unit cell with one atom in each plane. The unit cells used in the calculations are also shown.}
\end{center}
\end{figure*}
\begin{table}
\caption{\label{tab1}
Conductances per unit areas and TMR ratios calculated using supercells with 13\,ML of SrTiO$_3$. The units are in $\Omega^{-1}\mu{\rm m}^{-2}$ and $\%$, respectively. Here, $A=2.15\times10^{-7}\,\mu{\rm m}^2$ is the in-plane area of both the supercells.
}
\begin{ruledtabular}
\begin{tabular}{rcc}
\textrm{}&
\textrm{Co/SrTiO$_3$/Co(111)}&
\textrm{Ni/SrTiO$_3$/Ni(111)}\\
\colrule
$G_{{\rm P},\uparrow}/A$\,\,\,\,& 2.78$\times$10$^{-3}$ & 2.15$\times$10$^{-3}$\\
$G_{{\rm P},\downarrow}/A$\,\,\,\,& 1.51$\times$10$^{-1}$ & 3.56$\times$10$^{-2}$\\
$G_{{\rm AP},\uparrow}/A$\,\,\,\,& 1.20$\times$10$^{-2}$ & 4.85$\times$10$^{-3}$\\
$G_{{\rm AP},\downarrow}/A$\,\,\,\,& 1.22$\times$10$^{-2}$ & 4.84$\times$10$^{-3}$\\
$G_{\rm P}/A$\,\,\,\,& 1.53$\times$10$^{-1}$ & 3.78$\times$10$^{-2}$\\
$G_{\rm AP}/A$\,\,\,\,& 2.42$\times$10$^{-2}$ & 9.68$\times$10$^{-3}$\\
TMR ratio & 534 & 290\\
\end{tabular}
\end{ruledtabular}
\end{table}
Table \ref{tab1} shows the obtained conductances and TMR ratios in Co/SrTiO$_3$/Co(111) and Ni/SrTiO$_3$/Ni(111) MTJs. The Co-based MTJ exhibits a relatively high TMR ratio over 500\%, which is higher than that of the Ni-based MTJ (290\%). Note that the down-spin conductance $G_{{\rm P},\downarrow}$ is much larger than the up-spin conductance $G_{{\rm P},\uparrow}$ in both MTJs. This is a significant feature for the present TMR effect and its origin is discussed below.

In Fig. \ref{kpara}, we show the ${\bf k}_{\parallel}$-dependent conductances of the present MTJs, which provides key information to understand the mechanism of the TMR effect. Let us first focus on the Co/SrTiO$_3$/Co(111) MTJ with a higher TMR ratio [Figs. \ref{kpara}(a)--\ref{kpara}(c)]. In the down-spin conductance $G_{{\rm P},\downarrow}({\bf k}_{\parallel})$ in Fig. \ref{kpara}(b), one can see a smooth peak centered at ${\bf k}_{\parallel}=(0,0)=\Gamma$. This reminds us of the similar peak in Fe/MgO/Fe(001) \cite{2001Butler-PRB,2001Mathon-PRB}, which was explained by the coherent tunneling of the $\Delta_1$ state at ${\bf k}_{\parallel}=\Gamma$. In Fig. \ref{kpara}(a), the up-spin conductance $G_{{\rm P},\uparrow}({\bf k}_{\parallel})$ has a spikelike structure distributed circularly around the $\Gamma$ point. Such a feature is often seen in other MTJs and is known to come from the interfacial resonant tunneling. In the antiparallel magnetization state [Fig. \ref{kpara}(c)], the ${\bf k}_{\parallel}$ dependence of the conductance is like a mixture of $G_{{\rm P},\uparrow}({\bf k}_{\parallel})$ [Fig. \ref{kpara}(a)] and $G_{{\rm P},\downarrow}({\bf k}_{\parallel})$ [Fig. \ref{kpara}(b)] but the value of the conductance is small in each ${\bf k}_{\parallel}$ point, because of the mismatch of the conductive channels between the left and right electrodes. The ${\bf k}_{\parallel}$ dependences of conductances in the Ni-based MTJ [Figs. \ref{kpara}(d)-\ref{kpara}(f)] are almost similar to those of the Co-based MTJ. A minor difference is that the peak in the down-spin conductance [Fig. \ref{kpara}(e)] is not so smooth compared to that of the Co-based MTJ. However, the similarity in the ${\bf k}_{\parallel}$-dependent conductances indicates that the TMR effects in these MTJs can be explained by the same mechanism.

Let us discuss the mechanism of the TMR effect on the basis of the electronic structures of the tunnel barrier and ferromagnetic electrodes. We mainly focus on the Co/SrTiO$_3$/Co(111) MTJ, since the similar mechanism is expected for both the MTJs. As mentioned above, the smooth peak in the ${\bf k}_{\parallel}$-dependent conductance [Fig. \ref{kpara}(b)] reminds us of the well-known coherent tunneling mechanism, in which bulk band structures of the tunnel barrier and ferromagnetic electrodes along the $k_z$ line at the $\Gamma$ point can explain a high TMR ratio. In the conventional (001)-oriented MTJs, such a high symmetry line in the Brillouin zone is called the $\Delta$ line. On the other hand, in the present (111)-oriented MTJs, the $\Lambda$ line corresponding to the [111] direction plays the key role for the coherent tunneling.

Figure \ref{band-mixed}(a) shows real and complex band structures of SrTiO$_3$ along the $\Lambda$ line. Here, the Fermi level $E_{\rm F}$ is set to that of SrTiO$_3$ attached to Co, which was estimated by the Co/SrTiO$_3$/Co(111) supercell. Around $E=E_{\rm F}$, the real band has an insulating gap of $\sim 1.04\,{\rm eV}$, which is smaller than the typical theoretical value in SrTiO$_3$ ($\sim 1.9\,{\rm eV}$) estimated by similar first-principles calculations \cite{2003Tanaka-PRB,2015Kang-PRB}. This is because the in-plane lattice constant of SrTiO$_3$ is shrunk so as to fit that of fcc Co and the tensile strain ($\sim$23\%) is applied along the [111] direction. Because of such a small band gap in SrTiO$_3$, the present MTJs have small values of resistance-area product ($RA$), which are beneficial for realizing read sensors of high-density hard disk drives and Gbit-class magnetic random access memories. By calculating the inverse of $G_{\rm P}/A$ in Table \ref{tab1}, we obtained $RA$ of 6.52 and 26.46\,$\Omega\,\mu{\rm m}^2$ in the Co- and Ni-based MTJs, respectively. These values are much smaller than that in the typical Fe/MgO/Fe(001) MTJ with a similar barrier thickness ($\sim\, 10^3 \,\Omega\,\mu{\rm m}^2$) [see Fig. 4(b) of Ref. \cite{2017Masuda-JJAP}].

In Fig. \ref{band-mixed}(a), three complex bands cross $E=E_{\rm F}$, where one of them has $\Lambda_1$ symmetry (red curve) and the others have $\Lambda_3$ symmetry (black curves). The $s$, $p_z$, and $d_{3z^2-r^2}$ orbitals rotationally symmetric along the [111] direction belong to the $\Lambda_1$ state and the other $p$ and $d$ orbitals belong to the $\Lambda_3$ state. Note here that ${\rm Im}(k_z)$ provides a decay rate of the wave function in the barrier layer. Since all three complex bands have similar values of ${\rm Im}(k_z)$ at $E=E_{\rm F}$, both $\Lambda_1$ and $\Lambda_3$ wave functions of the electrode are expected to decay with a similar length scale in the SrTiO$_3$ barrier.

We next calculated the band structure of fcc Co along the $\Lambda$ line as shown in Figs. \ref{band-mixed}(b) and \ref{band-mixed}(c), using the unit cell extracted from the Co/SrTiO$_3$/Co(111) supercell. One can find the half-metallic nature in the $\Lambda_1$ state (red curves); namely, the $\Lambda_1$ band in the down-spin state crosses $E_{\rm F}$, while that in the up-spin state does not cross $E_{\rm F}$. The relatively high TMR ratio in this system is attributed to the coherent tunneling of the half-metallic $\Lambda_1$ state. However, the TMR ratio (534\%) is lower than that of the conventional Fe/MgO/Fe(001) MTJ ($>1000\%$) with a half-metallic $\Delta_1$ state in Fe \cite{2001Butler-PRB,2001Mathon-PRB}. As shown in Figs. \ref{band-mixed}(b) and \ref{band-mixed}(c), the $\Lambda_3$ band crosses $E_{\rm F}$ in both up- and down-spin states. Since the $\Lambda_3$ state has a similar decay rate as the $\Lambda_1$ state [Fig. \ref{band-mixed}(a)], these up- and down-spin $\Lambda_3$ bands enhance the conductance $G_{\rm AP}$ and thus decrease the TMR ratio. We also analyzed the band structure of fcc Ni in Ni/SrTiO$_3$/Ni(111) and found a similar half-metallic nature in the $\Lambda_1$ state (not shown). Thus, the TMR effect in the Ni-based MTJ can also be understood by the $\Lambda_1$ coherent tunneling, even though the TMR ratio is not so high compared to that of the Co-based MTJ. The origin of such a difference will be discussed later.

Note here that the half-metallic band structure in the $\Lambda_1$ state [Figs. \ref{band-mixed}(b) and \ref{band-mixed}(c)] can be interpreted as a result of the band folding in the $k_x$-$k_y$ plane. To discuss this, let us consider the primitive unit cell of (111)-oriented fcc Co shown on top of Figs. \ref{band-mixed}(d) and \ref{band-mixed}(e), which has one Co atom in each $ab$-plane layer and fits a simpler tunnel barrier with a smaller in-plane area like MgO. Using this unit cell, we calculated up- and down-spin band structures along the $\Lambda$ line shown in Figs. \ref{band-mixed}(d) and \ref{band-mixed}(e). It is seen that no band crosses $E_{\rm F}$ in both spin states, i.e., there is no $\Lambda_1$ state at $E_{\rm F}$. In the case of the larger unit cell that fits SrTiO$_3$ shown on top of Figs. \ref{band-mixed}(b) and \ref{band-mixed}(c), the $a$- and $b$-axis lengths are twice as long as those of the primitive cell and each $ab$-plane layer has four Co atoms. Therefore, the band structures of this extended cell are identical to those obtained by folding the band structures of the primitive cell in the $k_x$-$k_y$ plane. Actually, by comparing Figs. \ref{band-mixed}(b),\ref{band-mixed}(c) and \ref{band-mixed}(d),\ref{band-mixed}(e), we see that the band folding provides additional bands crossing $E_{\rm F}$, leading to the half metallicity in the $\Lambda_1$ state. We emphasize that this is in sharp contrast to the band-folding effect in Fe/MgAl$_2$O$_4$/Fe(001) \cite{2012Miura-PRB,2012Sukegawa-PRB,2017Masuda-PRB,2020Nawa-PRB,2021Nawa-PRApplied}; the band folding gives an additional minority-spin band in the $\Delta_1$ state of Fe, which breaks the $\Delta_1$ half metallicity and lowers a TMR ratio. In our previous study \cite{2020Masuda-PRB}, we studied the TMR effect in Co/MgO/Co(111). In this case of MgO(111), as mentioned above, the bulk band structure of Co has no $\Lambda_1$ state at $E_{\rm F}$ and cannot contribute to a high TMR ratio. Instead, in this system, we showed that the characteristic interfacial state gives a high TMR ratio through the resonant tunneling \cite{remark_CoMgOCo111}. However, such a high TMR ratio might be fragile against interfacial defects or impurities as mentioned in Sec. \ref{introduction}. In contrast, the presently obtained high TMR ratio is owing to the $\Lambda_1$ half-metallicity in the bulk electronic state and is expected to be more robust against interfacial imperfections than the interface-driven high TMR ratio.

\begin{table}
\caption{\label{tab2}
Conductances per unit areas and TMR ratios calculated for Co/SrTiO$_3$($n$\,ML)/Co(111) ($n=7, 13$). The units are in $\Omega^{-1}\mu{\rm m}^{-2}$ and $\%$, respectively.
}
\begin{ruledtabular}
\begin{tabular}{ccc}
\textrm{SrTiO$_3$ thickness}&
\textrm{7\,ML\,(11\AA)}&
\textrm{13\,ML\,(19\AA)}\\
\colrule
$G_{\rm P}/A$\,\,\,\,& 7.46 & 1.53$\times$10$^{-1}$\\
$G_{\rm AP}/A$\,\,\,\,& 2.20 & 2.42$\times$10$^{-2}$\\
TMR ratio & 240 & 534\\
\end{tabular}
\end{ruledtabular}
\end{table}
The above mentioned coherent tunneling scenario of the $\Lambda_1$ state is also supported by the SrTiO$_3$ thickness dependence of conductances and the TMR ratio. We additionally calculated these quantities in the Co/SrTiO$_3$(7\,ML)/Co(111) MTJ and compared them with those in the original Co/SrTiO$_3$(13\,ML)/Co(111) MTJ as shown in Table \ref{tab2}. We see that the TMR ratio increases with increasing the SrTiO$_3$ thickness. This is because the selective transport of the $\Lambda_1$ state becomes more prominent as the barrier thickness increases. A similar behavior is also seen in Fe/MgO/Fe(001) \cite{2001Butler-PRB,2001Mathon-PRB}, where the TMR effect originates from the selective transport of the $\Delta_1$ state. The decay of the parallel conductance $G_{\rm P}/A$ can be roughly estimated from the complex band shown in Fig. \ref{band-mixed}(a). When we use $\kappa={\rm Im}(k_z)=0.6\,\pi/c$ as the complex wave vector for the $\Lambda_1$ state, the decay factor of the conductance is calculated as $\exp(-2\kappa d)\approx 1.86 \times 10^{-2}$, where we used $d=8\,{\rm \AA}$ as the increment in the SrTiO$_3$ thickness (7$\rightarrow$13\,ML) and $c=7.57\,{\rm \AA}$ as the $c$-axis length of the SrTiO$_3$ cell. Using this factor and $G_{\rm P}/A$ for 7\,ML SrTiO$_3$, $G_{\rm P}/A$ for 13\,ML SrTiO$_3$ is approximately estimated as $G_{\rm P}/A\,(7\,{\rm ML}\,\,{\rm SrTiO_3}) \times \exp(-2\kappa d) \approx 1.39 \times 10^{-1}\,\Omega^{-1}\mu{\rm m}^{-2}$, which is close to $1.53 \times 10^{-6}\,\Omega^{-1}\mu{\rm m}^{-2}$ (Table \ref{tab2}) obtained in the actual transport calculation. All these results on the SrTiO$_3$ thickness dependence indicate that the coherent tunneling of the $\Lambda_1$ state driven by the band folding provides the dominant contribution to the TMR effect in the present systems.

\begin{figure}
\includegraphics[width=8.4cm]{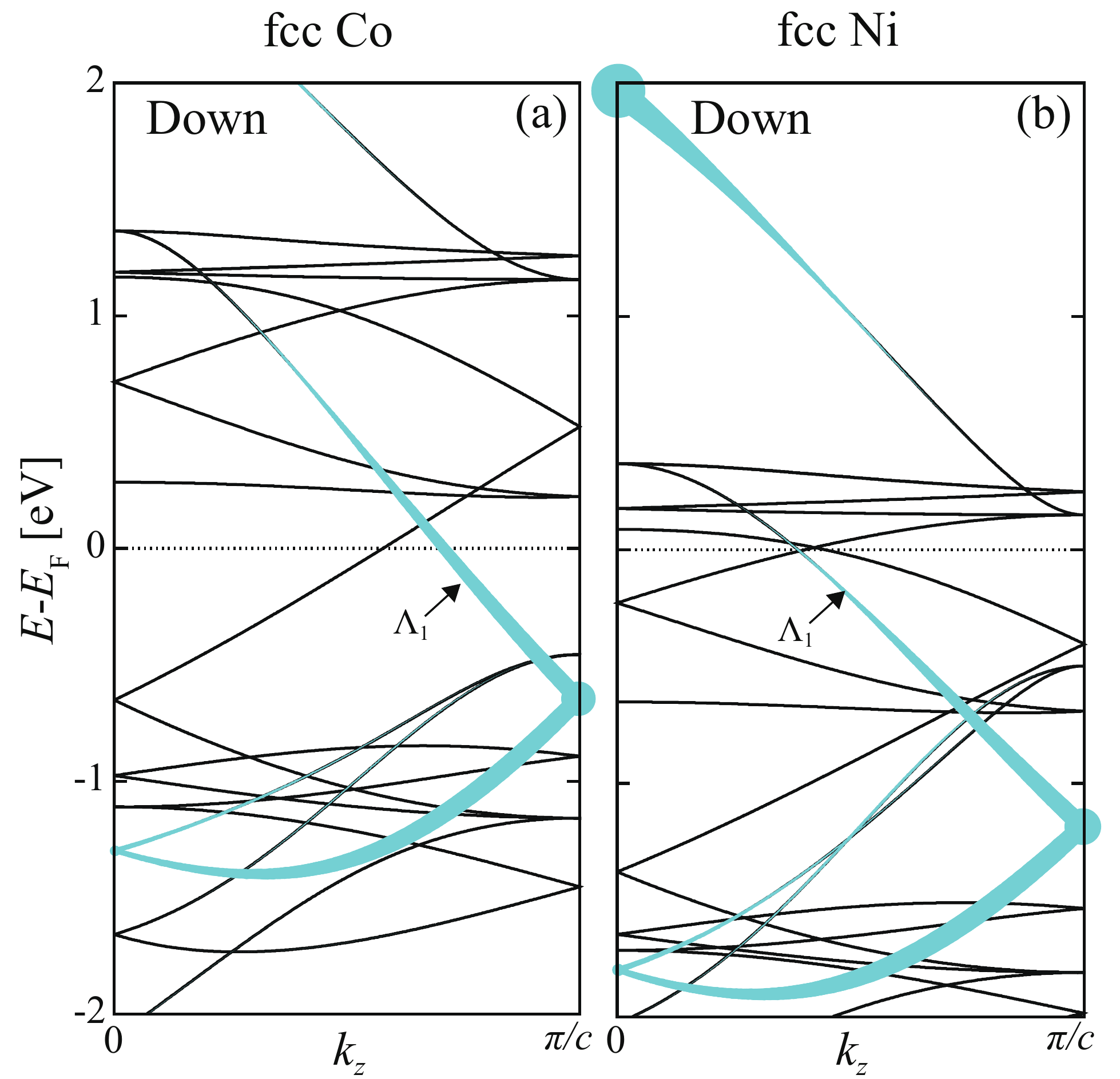}
\caption{\label{band-projection} Down-spin band structure along the $\Lambda$ line of (a) fcc Co and (b) fcc Ni with $s$-orbital projection. The relative weight of the $s$-orbital component is indicated as the linewidth of each band using color.}
\end{figure}

We finally address the difference in the TMR ratio between the Co- and Ni-based MTJs. As mentioned above, since both Co and Ni in the present MTJs have the $\Lambda_1$ half metallicity, a more detailed comparison of the electronic structure is required. Here, we focus on the weight of the $s$-orbital component in the $\Lambda_1$ band, since the $s$-orbital state contributes dominantly to transport properties including the TMR effect owing to its small effective mass. Such a significance of the $s$-orbital state on the TMR effect has been reported in many previous studies \cite{2014Li-PRB,2016Moges-PRB,2021Nawa-PRApplied,2021Masuda-PRBL}. Figures \ref{band-projection}(a) and \ref{band-projection}(b) show the down-spin band structures of Co and Ni, respectively, where the relative weight of the $s$-orbital component is indicated as the linewidth of each band using color. We find that the main $\Lambda_1$ band crossing $E_{\rm F}$ in Co has more $s$-orbital weight around $E_{\rm F}$ than in Ni. This larger $s$-orbital component at $E_{\rm F}$ can provide a larger conductance; in fact, as shown in Table \ref{tab1}, the down-spin conductance $G_{{\rm P},\downarrow}$ in the Co-based MTJ is more than four times larger than that in the Ni-based MTJ, while the up-spin conductance $G_{{\rm P},\uparrow}$ is almost similar for both MTJs. These results indicate that a higher TMR ratio in the Co-based MTJ is attributed to a larger $s$-orbital component at $E_{\rm F}$ in the $\Lambda_1$ band of Co.

\section{summary}
We investigated the TMR effect in unconventional (111)-oriented MTJs with SrTiO$_3$ tunnel barriers by means of the first-principles calculation and the Landauer formula. We obtained relatively high TMR ratios of 534 and 290\% in Co/SrTiO$_3$/Co(111) and Ni/SrTiO$_3$/Ni(111), respectively. The analysis of the bulk band structure in the electrode and the barrier regions of the MTJ clarified that the TMR effect in the present MTJs can be explained by the coherent tunneling of electronic states of bulk ferromagnets; actually, we found that fcc Co and Ni in the MTJs have half-metallic band structures in the $\Lambda_1$ state and these half-metallic states transmit through SrTiO$_3$ with an evanescent $\Lambda_1$ state, leading to relatively high TMR ratios. A usual primitive cell of fcc Co (Ni) has no $\Lambda_1$ state at $E_{\rm F}$. However, since the in-plane lattice constant of SrTiO$_3$ is about twice as long as that of fcc Co (Ni), the 2$\times$2 in-plane cell of fcc Co (Ni) fit the unit cell of SrTiO$_3$. This yields a band folding in fcc Co (Ni) in the $k_x$-$k_y$ plane and the folded bands give a half metallicity in the $\Lambda_1$ state. Therefore, we can conclude that the band folding is the key for the $\Lambda_1$ half-metallicity and resultant high TMR ratios. We also discussed the difference in the TMR ratio between the Co- and Ni-based MTJs and found that this is attributed to the different weights of the $s$-orbital component in the $\Lambda_1$ band at the Fermi level. Unfortunately, the TMR ratios of the present MTJs are not so high compared to that of the conventional Fe/MgO/Fe(001). This is because SrTiO$_3$ has a slow-decaying evanescent state with $\Lambda_3$ symmetry, as well as that with $\Lambda_1$ symmetry. Since fcc Co (Ni) has both up- and down-spin $\Lambda_3$ bands at $E_{\rm F}$, these bands degrade a TMR ratio by increasing the conductance in the antiparallel magnetization state. Therefore, finding ferromagnetic materials that have a half metallicity not only in the $\Lambda_1$ state but also in the $\Lambda_3$ state is required for obtaining a giant TMR ratio in the (111)-oriented MTJs with SrTiO$_3$ barriers.

\begin{acknowledgments}
This work was partly supported by Grants-in-Aid for Scientific Research (S) (Grant No. JP22H04966), Scientific Research (B) (Grants No. JP20H02190 and No. JP21H01750), and for Early-Career Scientists (Grant No. JP20K14782) from JSPS KAKENHI. This work was also supported by JST CREST ``Integrated Devices and Systems Utilized by Information Carriers'' (Grant No. JPMJCR21C1) and the Cooperative Research Project Program of the Research Institute of Electrical Communication, Tohoku University. The crystal structures were visualized using {\scriptsize VESTA} \cite{2011Momma_JAC}.
\end{acknowledgments}

% Create the reference section using BibTeX:
%\bibliography{basename of .bib file}

\begin{thebibliography}{99}
\bibitem{remark_barrier-thickness} Here, the barrier thickness is defined as the distance between two ferromagnetic layers closest to the SrTiO$_3$ barrier.
\end{thebibliography}

\end{document}